\documentclass[graybox]{svmult}
\pdfoutput=1

\usepackage{mathptmx}       
\usepackage{helvet}         
\usepackage{courier}        
\usepackage{makeidx}         
\usepackage{graphicx}        
\usepackage{multicol}        
\usepackage[bottom]{footmisc}
\usepackage{array}

\usepackage{subfigure}
\usepackage{pifont}

\makeindex             

\begin{document}

\title*{Visual Exploration of Simulated and Measured Blood Flow}
\titlerunning{4D Flow}
\author{ A. Vilanova\inst{1}, Bernhard Preim\inst{2} \and Roy van Pelt\inst{1} \and \\
Rocco Gasteiger\inst{2}
\and Mathias Neugebauer\inst{2} \and Thomas Wischgoll\inst{3} }%

\institute{
1 Dept. of Biomedical Engineering, Eindhoven Univ. of Technology, The Netherlands\\
email: \{a.vilanova, r.f.p.v.pelt\}@tue.nl, \\
2 Institut f\"ur Simulation und Graphik, Otto-von-Guericke-Universit\"at Magdeburg \\
email: {bernhard.preim, rocco.gasteiger, mathias.neugebauer}@ovgu.de \\
3 Computer Science and Engineering, Wright State University \\
email: thomas.wischgoll@wright.edu, }

\authorrunning{Vilanova et al.}
%
%
\maketitle

\abstract{ Morphology of cardiovascular tissue is influenced by the unsteady behavior of the blood flow and vice versa. Therefore, the pathogenesis of several cardiovascular diseases is directly affected by the blood-flow dynamics. Understanding flow behavior is of vital importance to understand the cardiovascular system and potentially harbors a considerable value for both diagnosis and risk assessment.
The analysis of hemodynamic characteristics involves qualitative and quantitative
inspection of the blood-flow field. Visualization plays an important role in the qualitative
exploration, as well as the definition of relevant quantitative measures and its
validation.
There are two main approaches to obtain information about the blood flow: simulation
by computational fluid dynamics, and in-vivo measurements. Although research
on blood flow simulation has been performed for decades, many open problems remain
concerning accuracy and patient-specific solutions. Possibilities for real measurement
of blood flow have recently increased considerably by new developments
in magnetic resonance imaging which enable the acquisition of 3D quantitative measurements
of blood-flow velocity fields.
This chapter presents the visualization challenges for both simulation and real
measurements of unsteady blood-flow fields. 
}


\section{Introduction}\label{section-intro}

Cardiovascular disease (CVD) is a class of conditions affecting the heart and blood vessels, with an estimated overall prevalence of over thirty percent of the American population~\cite{AmericanHeart:2010}, and is currently the leading cause of death worldwide~\cite{WorldHealthOrganization:2011}.

Diagnosis of CVD typically involves an evaluation of both the anatomical structure and function, while the behavior of blood flow is still rarely inspected. The flow behavior is, nevertheless, of vital importance to the cardiovascular system. Morphology of cardiovascular tissue is significantly influenced by the unsteady behavior of flowing blood and vice versa. Therefore, blood flow analysis potentially harbors a considerable value for both diagnosis and risk assessment. A wide range of pre-clinical research indicates that flow behavior directly relates to medical conditions~\cite{Gatehouse:2005,Morbiducci:2009}.

In particular, congenital heart diseases imply anomalous hemodynamics that strongly influence the progression and treatment of the innate defects. For the adult case, a noteworthy application is the aortic dissection, which is caused by a tear in the inner aortic wall. This allows blood to flow between the disintegrated layers of the vessel wall, resulting in a high risk of rupture. Again, the blood flow behavior plays a predominant role in the course of the condition. Decision support in case of cerebral aneurysms is one of the main applications of blood flow analysis. Blood flow is essential for the assessment of risk of rupture, urgency of treatment in case of multiple aneurysms, selection of treatment strategy (e.g., coiling/stenting, neurosurgical clipping).

The analysis of hemodynamic characteristics involves qualitative and quantitative inspection of the blood flow field. Physicians in clinical research investigate both the spatiotemporal flow behavior, as well as derived measures, such as the mean flux or cardiac output. The analysis of the blood flow data often requires complex mental reconstruction processes by the physician. Visualization plays an important role in the qualitative exploration, as well as the definition of relevant quantitative measures and its validation.

There are two main approaches to obtain information about the blood flow: simulations (i.e., computational fluid dynamics) and in-vivo measurements. Both of these methodologies can obtain information about the unsteady blood flow characteristics, where each has different advantages and disadvantages.

Although research on simulations of blood flow has been active for several decades, still a lot of open problems remain concerning accuracy and patient-specific solutions. Recently, research around measurement of blood flow has increased considerably. Developments in magnetic resonance imaging (MRI) have made the acquisition of 3D quantitative measurements of blood flow velocities fields possible.
Furthermore, several vendors have made essential postprocessing software to inspect the data clinically available. Therefore, clinical pilot studies have been possible and have shown the relevance and potential of this data. Furthermore, the MRI acquisition development towards 7 and 9 Tesla machines have the potential to provide the required resolution and signal-to-noise ratios (SNR) to analyze blood flow in smaller vessels compared to the main arteries around the heart.

In this chapter, we will consider the visualization challenges for both simulations and in-vivo measurements of unsteady flow fields. We will present a review of the existing literature, the main challenges related to blood flow visualization and analysis, as well as the open issues.

\section{Blood Flow Simulation}\label{section-simulation}


One way of determining blood flow within a vascular system is through simulation. This approach involves two major steps. First, the
vascular structure needs to be segmented, and the geometry of the
vessel boundary determined as accurately as possible. Next, a
Computational Fluid Dynamics (CFD) model simulates the blood flow
within the reconstructed geometry. The next sections will discuss these steps in more
detail.

\subsection{Grid Generation}

In order to simulate blood flow with CFD, the boundary conditions of the underlying mathematical model have to be defined properly. In case of vascular flow, the boundary conditions
are defined by two different components: the first one is the the
geometric boundary of the vessels; the second one consists of the inflow and
outflow characteristics as defined by the circulatory system.

There are different ways of identifying the vessel
boundary. Typically, some imaging technique is used to generate a scan
of the vascular structure for which the flow is supposed to be
simulated, for example a Computed Tomography (CT) scan. In
order to extract the vascular structure from such a volumetric image,
the data needs to be segmented. Simple thresholding based on the
intensity value can be used. However, this may not be sufficient for
anatomical structures where significant perfusion and noise occurs, such as the
heart. More sophisticated segmentation techniques are necessary, for example, gradient-based thresholding techniques tend to produce better results in those cases.

The segmentation process also has great influence on the overall
accuracy of the simulation. Basic intensity thresholding techniques,
for example, determine individual voxel locations as being
part of the vessel boundary. However, it is unlikely that the vessel
boundary is located precisely at such a voxel location, especially
given that the volumetric data set only imposes an
artificial grid on the organ at hand. Therefore, the accuracy can be
improved by using segmentation techniques that operate at a sub-voxel
level, e.g., Marching Cubes algorithm~\cite{Lorensen:1987}.

\begin{figure}[h]
\centering
\subfigure[]{\includegraphics[height=40mm]{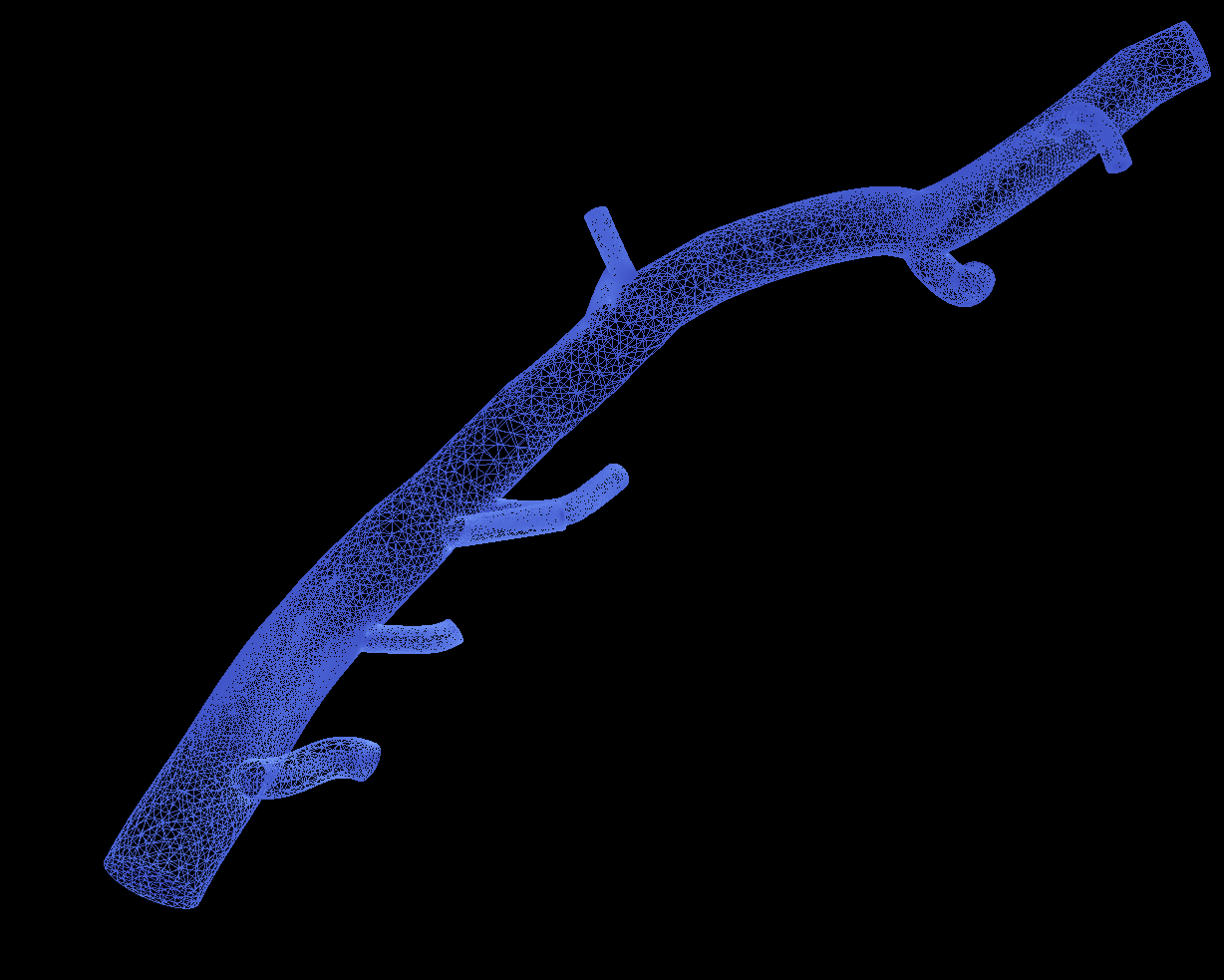}}\quad
\subfigure[]{\includegraphics[height=40mm]{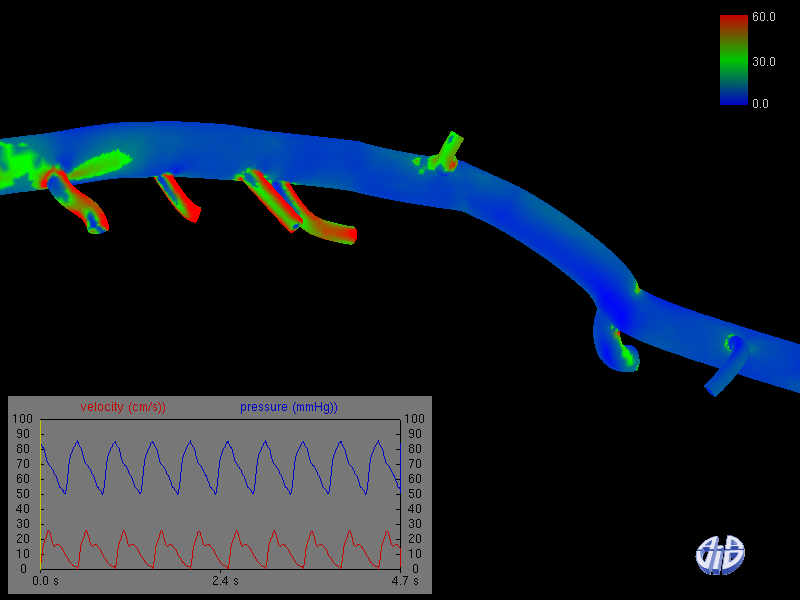}}
\caption{An example of a vessel boundary based on a CT scan segmented using
gradient-based thresholding with sub-voxel precision (a) and resulting
simulation of blood flow visualized by color-coding the
wall-shear stress (b).}
\label{fig:simulation}
\end{figure}

The downside of the Marching Cubes algorithm,
however, is that it only operates with a fixed intensity threshold
across the entire data set. Since typically a point
spread function with a radius greater than one has to be assumed for
most imaging techniques, this may cause errors in the boundary
geometry, overestimating larger vessels and underestimating
smaller vessels.
Gradient-based
approaches can achieve better results in these cases
 identifying the location where the maximal gradient value is assumed
to find a more accurate estimate for the exact location of the vessel
boundary~\cite{Wischgoll:2008}. The geometric model resulting from the
segmentation step can then be further refined, for example by using
smoothing or rounding off the transitions at vessel
bifurcations~\cite{Oeltze:2005}, resulting in a vessel boundary that
can be used for a CFD simulation. Figure~\ref{fig:simulation}(a) shows an example of such a
vessel boundary generated based on a CT scan using gradient-based thresholding
with sub-voxel precision.

\subsection{Computational Fluid Dynamics Model}

In addition to the geometric boundary of the vascular structure
inflow, outflow, and wall boundary conditions have to be defined
properly for the CFD
simulation~\cite{Cebral2005,Venugopal2007}. Inflow and outflow
conditions arise from the fact that the current vascular structure has
to be isolated from the rest of the arterial system. In practice, the
flow rate or the speed profile at the inlets and the pressure at the
outlets are utilized for a cardiac cycle. These quantities are
obtained based on experimental measurements or by time-resolved
Phase-contrast MRI flow measuring from the patient (see subsequent
section).  boundary condition arises from the fact that the vessel
wall is distensible, which may influence the local hemodynamic and
vice versa~\cite{Torii2009}. However, typically no proper
characterization of the arterial wall, such as modulus of elasticity,
wall thickness, or pressure wave that form at the wall, is available
or is difficult to measure noninvasively~\cite{Cebral2005}. Thus, a
rigid wall is assumed in most cases, which also decreases the
numerical computation time. With all boundary conditions defined, the
flow can be computed based on the 3D unsteady Navier-Stokes equations
for an incompressible Newtonian fluid. Typically, common CFD solvers
are used for this step, for example ANSYS Fluent or OpenFOAM. The
resulting velocity and pressure of the blood flow can subsequently be
used for further analysis. For example, the computation of the
dimensionless Reynolds number~\cite{Cebral2009}. The Reynolds (Re)
number characterizes the local flow behaviour in terms of laminar (Re
$< 300$) or turbulent (Re $> 300$). In addition to velocity and
pressure values, other hemodynamic quantities are obtained during the
simulation. An important quantity is the wall shear stress
(\emph{WSS}), which represents the tangential force produced by blood
moving across the vessel surface. It is known that \emph{WSS} has an
influence on the tissue structure of the vessel wall and it is likely
that \emph{WSS} plays an important role in initiation, growth and
rupture of cerebral aneurysms~\cite{Nixon2010}. The \emph{WSS} can be
computed based on the velocity field and the
geometry~\cite{Huo:2009,Cebral2009}. Figure~\ref{fig:simulation}(b)
shows the result of such a CFD simulation using the inlet pressure and
velocity based on a typical heart rate. This simulation is based on
124 time steps. For each time step, a grid size of 500,000 cells was
used to accurately represent the vascular structure resulting in close
to 900MB of data. CFD simulations give blood flow information at high
resolution. However, CFD simulations are based on models with
assumptions and simplifications which make it difficult to obtain
patient-specific accurate results.

\section{Blood Flow Measurement}\label{section-measureddata}

\subsection{Acquisition Methods}

Measured blood flow information is mostly obtained by quantitative ultrasound (US) acquisition (see chapter 5). US is a cost-effective modality, providing flow information at high spatiotemporal resolution. However, US acquisition requires a skilled operator, is generally subject to a substantial amount of noise, and volumetric measurements of the vector velocity field are not possible. Consequently, this modality is less suitable for challenging cardiovascular conditions. Alternatively, computed tomography provides a limited number of blood flow acquisition sequences while delivering better signal-to-noise ratios. CT has the drawback of not measuring flow directly and exposing the patient to harmful radiation, which is impermissible for young patients. Instead, we focus on non-invasive Phase-Contrast (PC) MRI acquisition, which is the only modality providing volumetric quantitative measurements of blood flow velocities throughout the cardiac cycle. A typical size of such a volumetric data is $150\times150\times50$ voxels with velocity vectors with a resolution of $2\times2\times2.5$mm per voxel, and a time series of 20 to 25 steps per cardiac cycle.

\begin{figure}[h]
\centering
\subfigure[]{\includegraphics[height=37mm]{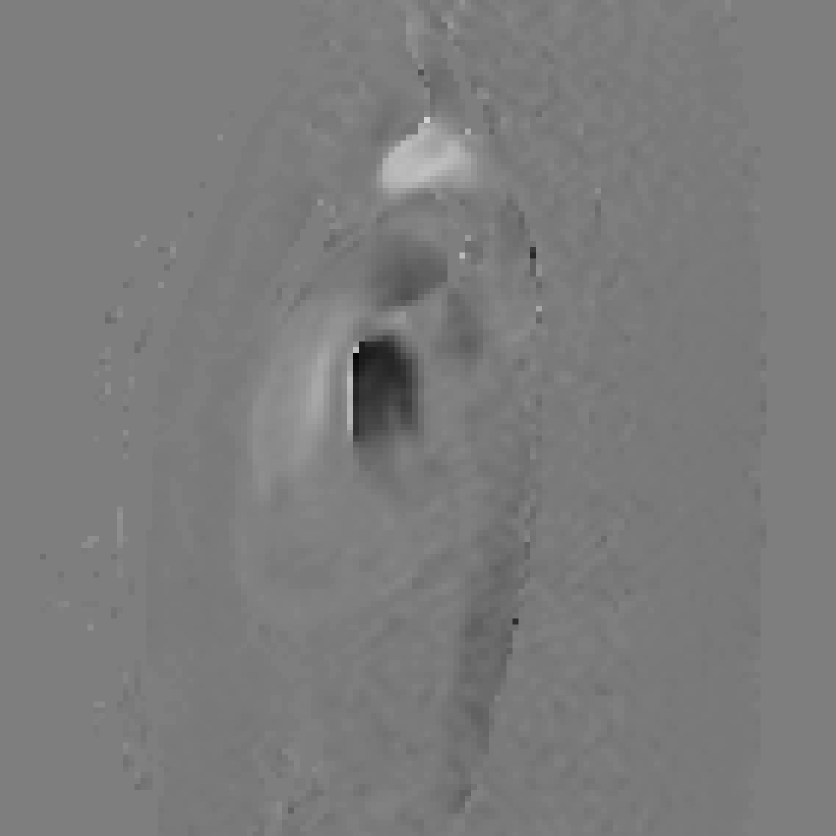}}\quad
\subfigure[]{\includegraphics[height=37mm]{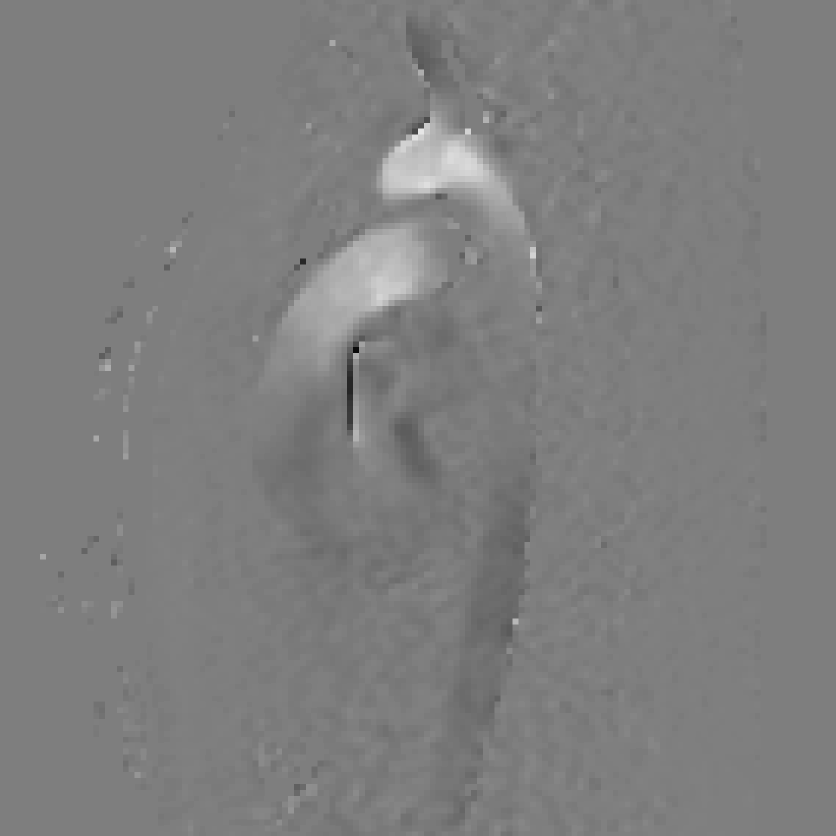}}\quad
\subfigure[]{\includegraphics[height=37mm]{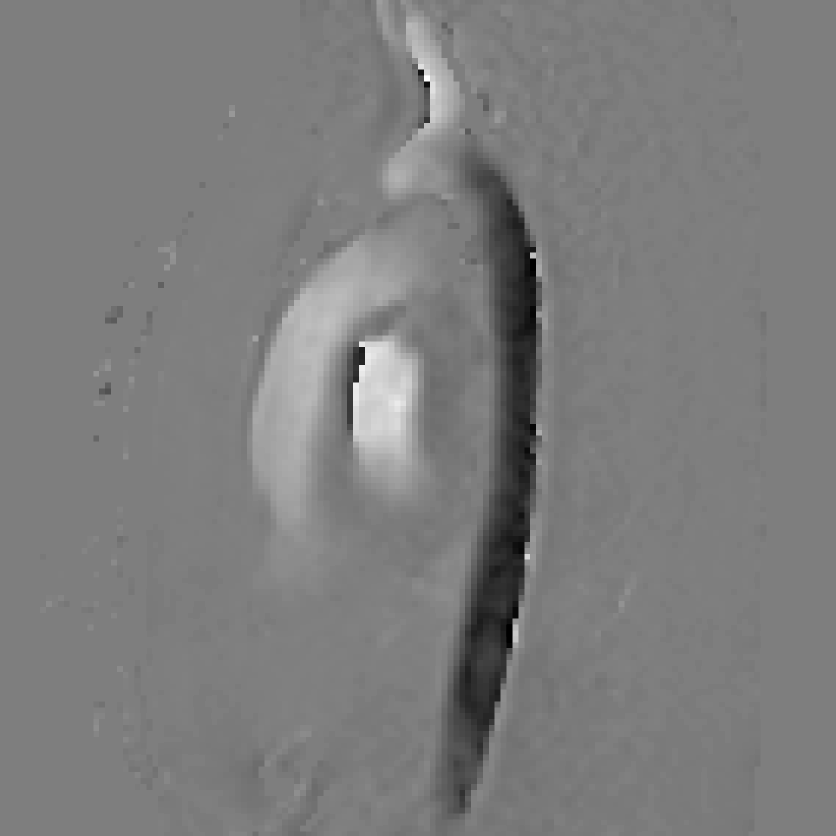}}\\
\subfigure[]{\includegraphics[height=37mm]{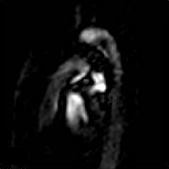}}\quad
\subfigure[]{\includegraphics[height=37mm]{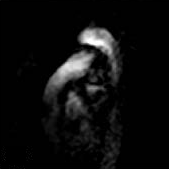}}\quad
\subfigure[]{\includegraphics[height=37mm]{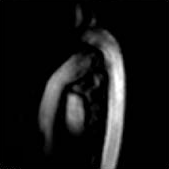}}
 \caption{The PC flow data set consists of 20 phases in time. Each phase in the series comprises a velocity vector volume with a resolution of $144\times144\times50$ voxels of $2.0\times2.0\times2.7$ mm. (a) PC-P right to left (b) PC-P anterior to posterior (c) PC-P head to feet (d) PC-M right to left (e) PC-M anterior to posterior (f) PC-M head to feet.}
 \label{fig:PCAdata}
\end{figure}

Phase-contrast MRI sequences enable acquisition of flow data that is linearly related to the actual blood flow velocities, capturing both speed and direction. This linear relation is described by the velocity encoding (VENC) acquisition parameter, representing the largest speed that can be measured unambiguously and is typically defined in centimeters per second. The range of the imposed speed limit, for example (-100 cm/s, 100 cm/s], corresponds to the phase extremities, i.e., $-\pi$ and $\pi$ radians. If a suitable VENC is chosen, PC-MRI provides a data set with great correspondence to the actual blood flow velocity field~\cite{greil:2002}. As a consequence, the acquired data allows for quantitative analysis of the blood flow behavior. PC cine MRI sequences support the acquisition of volumetric blood flow data throughout the cardiac cycle, generating a 4D blood flow velocity field~\cite{pelc:1991,markl:2003}. There are two customary approaches to reconstruct the acquired raw data to the desired flow images~\cite{bernstein:1991}: phase (PC-P) and magnitude (PC-M) reconstruction. Figure~\ref{fig:PCAdata} depicts a single slice of the reconstructed 4D flow data, at a certain point in time. The top row, Figures~\ref{fig:PCAdata}(a) - (c), represents the blood flow data in the three patient-oriented orthogonal directions, encoding both speed and directions of the blood flow quantitatively. This data is commonly referred to as the phase (PC-P) reconstruction. The bottom row, Figures~\ref{fig:PCAdata} (d) - (f), represents the blood flow data in three directions, encoding only speed. This data is commonly referred to as the complex difference or magnitude (PC-M) reconstruction. Even though the blood flow direction cannot be resolved from the PC-M reconstruction, the resulting data is inherently less prone to the uncorrelated noise that is typical for the PC-P reconstructed data.


\subsection{Noise and Artifacts}
Measurements come with imperfections which complicate the meaningful quantitative analysis of the flow data. In particular, measures derived from the data are sensitive to relatively small errors in the flow measurements. Inaccuracies are caused by a combination of many factors, associated with the MRI hardware, imaging sequences and their parametrization, and patient movement. For sensitive cardiac applications, the generally accepted objective is to acquire flow data with less than 10\% error~\cite{Gatehouse:2010}.

The parametrization of the imaging sequence has a large influence on the accuracy of flow measurements. The parametrization directly influences the spatial and temporal resolution~\cite{greil:2002}. In particular quantitative analysis of small vessels (e.g., in the brain) becomes cumbersome at low resolution~\cite{arheden:2001}.

Besides user parametrization, motion is an important cause of imaging artifacts. There are three major causes of tissue displacement due to patient movement: motion artifacts by peristaltic motion, artifacts caused by contraction of the heart muscle, and respiration. Contraction of the heart muscle artifacts can be considerably reduced during acquisition. The impact of the respiratory motion can also be largely suppressed, by exploiting the relatively motionless period after exhalation.

In addition, flow measurements are subject to general MRI artifacts, largely due to hardware imperfections common in all MRI scans~\cite{Bernstein:1998}. A relevant artifact for flow is due to the fast gradient switching which induces eddy currents in the electromagnetic field. This causes background phase errors in the image, which manifest as slowly varying image gradients in both the spatial and temporal domains. These effects are difficult to predict and therefore challenging to correct~\cite{Gatehouse:2010,Rolf:2011}. The conventional MRI noise follows a Rician distribution. For flow imaging, it can be shown that the noise in flow regions depends on the velocity encoding speed and is inversely proportional to the SNR of the corresponding magnitude image~\cite{lotz:2005}. Hence, the VENC parameter should be chosen as small as possible, while capturing the full dynamic range of the actual flow. The decision about the VENC value is often not easy to make.

There are additional artifacts that are specific to flow data. For instance, aliasing, or phase wrapping, erroneously introduces regions with opposite flow directions. Whenever the actual blood flow speed transcends the VENC value, a \emph{phase wrap} occurs. Several methods have been devised to correct these artifacts caused by a single phase wrap through postprocessing~\cite{yang:1996,langley:2009}.
Another flow-specific artifact is misregistration where blood flow regions are shifted from the stationary tissue. This is due to the time between the phase encoding and frequency encoding gradients. These artifacts can be corrected by adding a bipolar gradient to each phase encoding gradient~\cite{vlaardingerbroek:1999}.

The flow imaging sequences are based on the assumption that the blood flow velocities are constant at the time of measurement. Hence, measurements of accelerated flows are less accurate and can cause undesirable artifacts. Accelerated flows can be found in pulsatile flows, stenotic flows, or jets that can cause so-called flow voids at relatively low spatial resolutions. To verify whether this behavior has occurred, black blood scans are often employed to inspect the vessel delineation. 

\section{Visual Exploration}\label{section-visualexploration}

\subsection{Visualization of the Anatomical Context}\label{section-anatomicalcontext}
\begin{figure}[h]
\centering
 \includegraphics[width=0.8\textwidth]{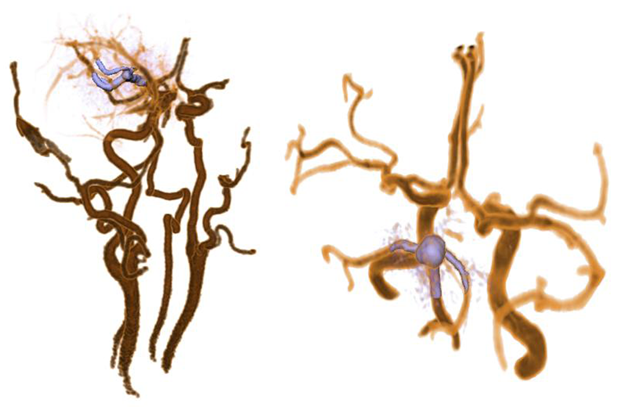}
\caption{The focus structures, the aneurysm and its immediate inflow and outflow are rendered opaquely with highly saturated colors. Context information is shown with a less striking colour with decreasing opacity for more distant vasculars structures.}~\label{fig:focus_context}
\end{figure}

The visual exploration of blood flow data is usually
focused on a rather small anatomical region. In case of simulated
blood flow data, this represents the domain where the simulation
was performed. It may be necessary to present this focus region
embedded in a somehow larger context to better understand the
location of a pathology and the in- and outflow regions. Such a
visualization goal may be achieved with a coordinated
focus-and-context view, where the detail view presents only the
target region and the context view provides the big picture with
additional anatomical context. An integrated
focus-and-context view is mentally easier to interpret. A reasonable
strategy is to employ distance-based transfer functions
\cite{Tappenbeck_2006_SimVis}, where the distance to the target
anatomy is mapped to opacity in order to hide distant vascular
structures. This strategy is illustrated in Figure~\ref{fig:focus_context}. The specific choice of colours and opacity
as well as the amount of information to be displayed requires
careful discussions with physicians~\cite{Neugebauer_2009_CARS}. Such a
visualization may be a first step in a pipeline of exploration and
analysis, as it presents an overview and needs to be
followed by a more local analysis.

The flow strongly depends on local variations of the enclosing
vascular structures. Large changes in flow speed occur at stenotic
regions, and turbulent flow occurs primarily at bifurcations or
strongly curved areas. Thus, it is important to investigate the
morphology of anatomical structures and the internal flow
simultaneously. In case of simulated flow, such an integrated
analysis may reveal that a significant flow feature is due to a
small variation of the surface, which may result from an
inaccuracy in the segmentation. The simplest idea to
display flow and vascular anatomy at the same time is to render the
vascular surface transparently. However, depending on the
transparency level, either the vascular anatomy is hardly
recognizable, or the internal flow is strongly obscured by the
vessel wall.

\begin{figure}[t]
\centering
\subfigure[]{\includegraphics[width=0.3\textwidth]{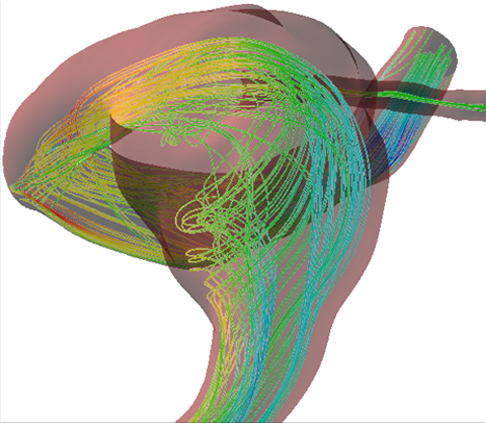}}\quad
\subfigure[]{\includegraphics[width=0.3\textwidth]{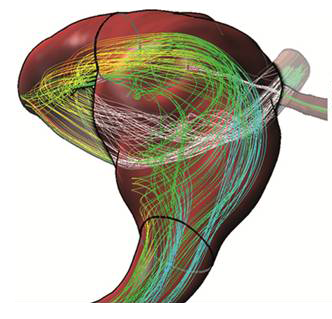}}\quad
\subfigure[]{\includegraphics[width=0.3\textwidth]{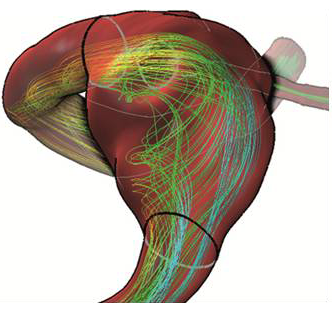}}
 \caption{Comparison between a semi-transparent visualization (a) and a ghosted view technique (b and c) applied to the enclosed vessel surface to show the internal flow. In (b) the hidden streamlines are depicted in grey and omitted in (c) to reduce visual clutter}
 \label{fig:ghostedview2}
\end{figure}

As a remedy, smart visibility techniques \cite{Viola:2005}, such as
ghosted views, may be employed. The flow may be considered as an important object and the vessel
walls transparency is modified to reveal flow lines. This idea has been realized
by Gasteiger et al.~\cite{Gasteiger_2010_VCBM}. The specific
solution to provide ghosted view visualizations is based on a
Fresnel reflection model~\cite{Schlick:1993}, where the reflection term is replaced by opacity. In Figure~\ref{fig:ghostedview2} a
comparison of that technique with conventional semi-transparent
rendering is presented. Gasteiger et al. refined their technique by an integration of landmarks described in the next section, and the ability to remove all hidden flow lines to further
reduce visual clutter (see Figure~\ref{fig:ghostedview2}).

\begin{figure}[h]
\centering
\subfigure[]{\includegraphics[height=37mm]{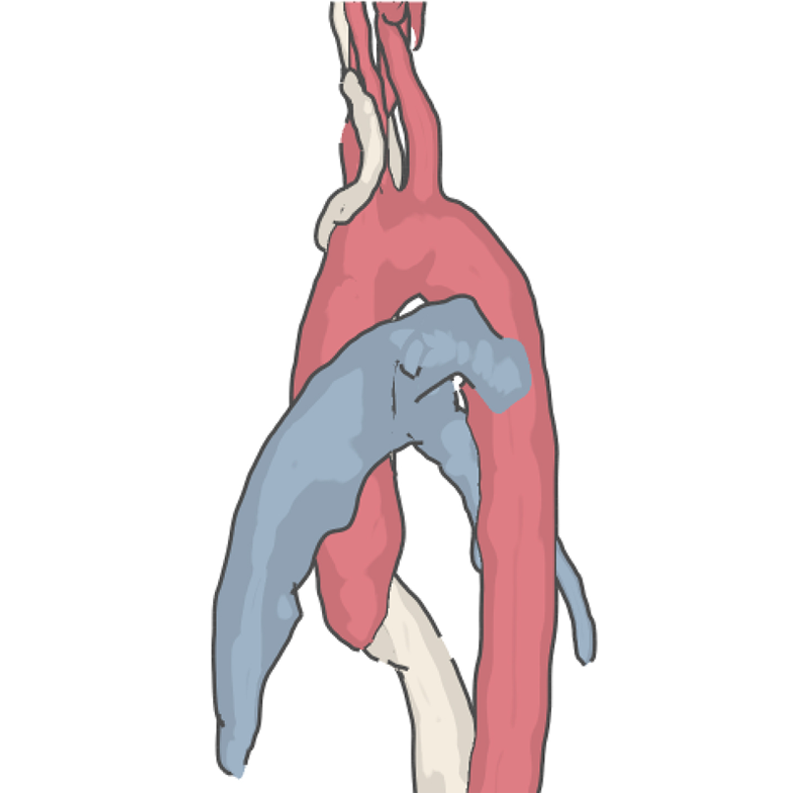}}\quad
\subfigure[]{\includegraphics[height=37mm]{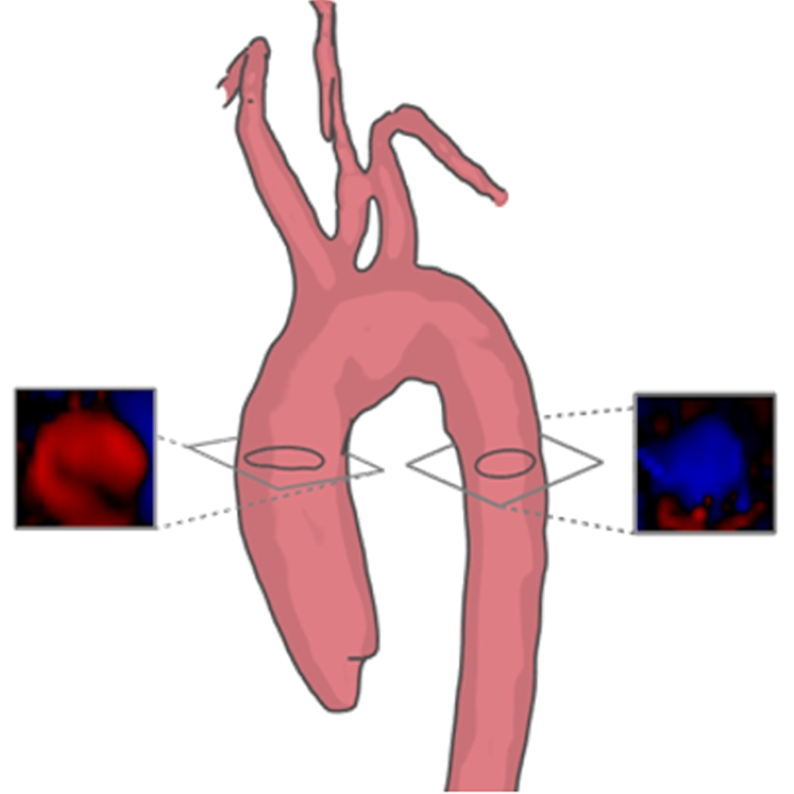}}\quad
\subfigure[]{\includegraphics[height=37mm]{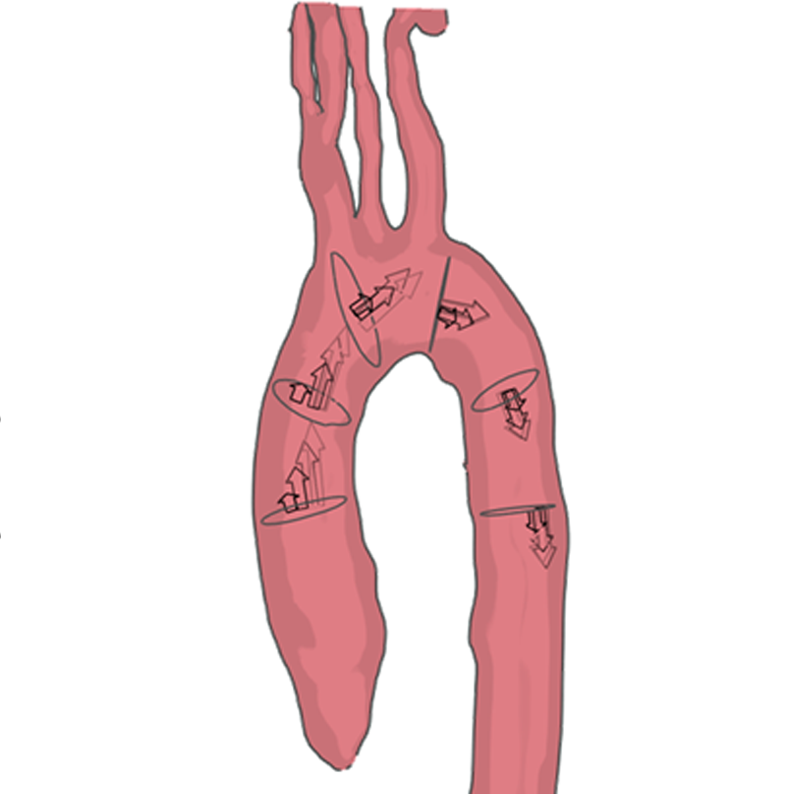}}
 \caption{Anatomical context visualization with cel shaded silhouettes and occluding contours. (a) Thoracic arteries. (b) MPR visualization exploded view of planes at cross-section positions. (c) Flow-rate arrow-trails arrows at 280 ms after the start of the cardiac cycle.~\cite{Pelt:2010}}
 \label{fig:vanPelt2010:context}
\end{figure}

Van Pelt et al.~\cite{Pelt:2010} presented an anatomical context based on methods inspired by medical illustrations, where the detail is removed while the morphological information is preserved(see Figure~\ref{fig:vanPelt2010:context}). To this end, they used cel-shaded silhouettes, combined with superimposed occluding contours. Hidden contours were visible during viewpoint interactions in order to resolve occlusion problems and to clarify spatial relations. Their user evaluation showed that these methods had a positive impact for the purpose of anatomical context representation of the flow.


\subsection{Localization of Anatomical Landmarks}

The exploration of vascular structures and the embedded flow
benefits strongly from geometric descriptors that enable a carefully
guided or constrained interaction. Such geometric descriptors may
also be used to decompose the relevant anatomy in meaningful
subregions to ease the exploration of complex flow data.

A widely used geometric descriptor is the vessel centerline,
determined by a skeletonization algorithm (see, e.g.,
\cite{Kirbas:2004}). The vessel centerline is often used in order to
move a cross-sectional plane that is always aligned perpendicular to
the centerline, presenting the maximum-sized area. In conventional
vessel analysis packages, the cross-sectional view displays the
intensity values from the original image data, e.g., the CT
Hounsfield values. In case of blood flow data, this strategy may be
used to present any scalar value derived from the flow or the flow
data itself, e.g., by using some glyph mapping. Van Pelt et al.~\cite{Pelt:2010} presented this cross-sectional visualization approach for the main arteries (see Figure~\ref{fig:vanPelt2010:context}).

\begin{figure}[h]
\centering
  \includegraphics[width=0.5\textwidth]{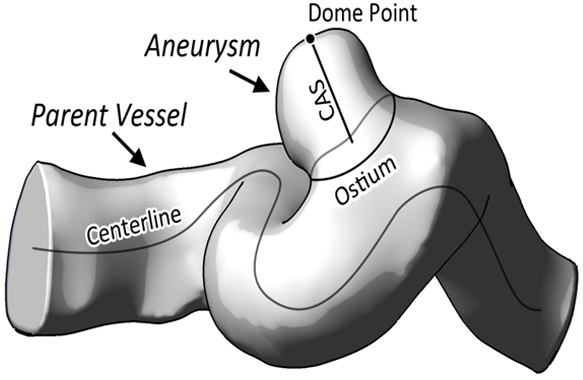}
\caption{Geometric descriptors for characterizing cerebral aneurysms
(From: Neigebauer et al.~\cite{Neugebauer_2010_VMV})}~\label{fig:landmarks}
\end{figure}

Better support for exploration tasks may be achieved by detecting
and analyzing further anatomical landmarks of a particular region.
Once these landmarks are identified, they may be used for labeling
and for guiding movements in the complex 3D anatomy. The choice of such landmarks is specific for a particular anatomical region. We describe
and illustrate this principle for the specific example of cerebral
aneurysms. First, it is essential to understand \emph{which}
landmarks are actually important to characterize the local vessel
anatomy. Neugebauer et al. \cite{Neugebauer_2010_VMV} questioned a couple
of neuroradiologists to draw cerebral aneurysms and extracted
characteristic points commonly used by them. The following points were deemed essential (see Figure~\ref{fig:landmarks}).
\begin{itemize}
\item the \emph{dome point} of an aneurysm,
\item the (curved) \emph{ostium plane}, where the blood enters the aneurysm, and
\item a so-called \emph{central aneurysm axis} (the closest connection between the parent vessel's centerline and the dome point)
\end{itemize}

 These landmarks may be utilized for example to
 move a plane along the central aneurysm axis or to employ the ostium plane
 as the seeding region, where streamline integration starts. The robust
 and precise detection is challenging due to the large variety of pathologic
 situations. Nevertheless even if it is successful only in about 90 percent of the
 cases, it provides a valuable support (see Neugebauer et al.~\cite{Neugebauer_2010_VMV}
 for a description of landmark extraction in saccular cerebral aneurysms).
 A similar landmark extraction process can be significant for
 other anatomical regions as well, since it provides a familiar reference frame for
 medical doctors. Constrained navigation does not necessarily mean that
  it is impossible to deviate from a predefined path. There
 are many variants to combine a preference direction with free exploration
 where the user is attracted to the predefined path, but may deviate.
 For a general discussion of constrained navigation techniques, see Hanson et al.\cite{Hanson:1997}
 and more recently Elmqvist and Tudoreanu~\cite{Elmqvist:2008}.


\subsection{Exploration of Surface Flow Scalar Features}

Blood flow simulations result in flow data as well as scalar flow features,
such as pressure, speed and wall shear stress (WSS). It is known that WSS plays an essential role in the understanding of
initiation and progression of vascular diseases. A simple solution
to display scalar flow features is to show the surface of the
relevant vascular region with a color-coded scalar flow feature.

The disadvantage of this simple solution is that only a small
portion of the surface is visible at the same time. Map projections,
which unfold an anatomical structure onto a plane, allow the visualization of the
whole scalar information simultaneously. However, a map exhibits
distortions (not all spatial relations, such as distances or angles
can be preserved) and, even worse, a simple map is very hard to
relate to the complex 3D anatomy of pathologic vessels. One promising approach is to combine a faithful 3D anatomy representation and a map view,
where interaction in both views are synchronized. It is
inspired by map views in other areas of medical diagnosis, such as
the Bull's eye plot in cardiology and stretched curved planar
reformations in vessel diagnosis \cite{Kanitsar:2002}.

\begin{figure}[h]
\centering
\includegraphics[width=.48\linewidth]{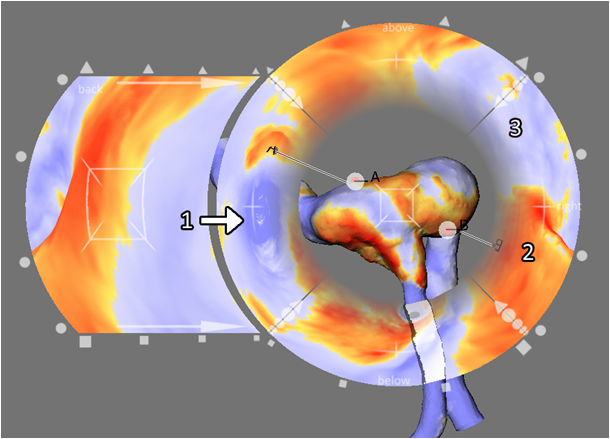}
\includegraphics[width=.48\linewidth]{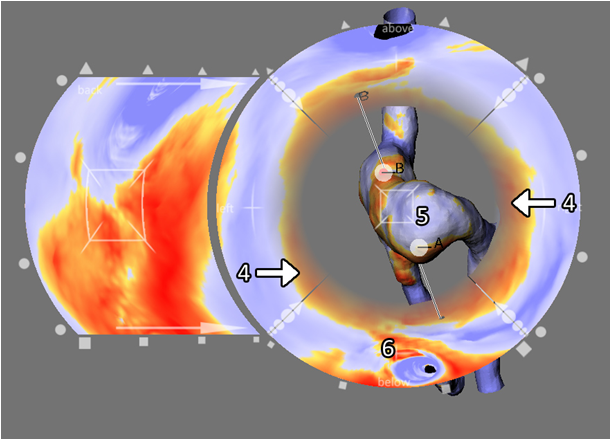}
\caption{A 3D model of the relevant vascular anatomy is surrounded
by map views that display scalar flow features of five sides
(features at the left, right, bottom, and up side are shown at the
corresponding ring portions). Scalar features of the backside are
shown at the most right display. The lines pointing from the map
portions to the 3D view indicate correspondences, where scalar
features are shown in both views. If the user drags a point,
representing an interesting feature from a map view to the center,
the anatomical model is rotated to make that region visible. All map
views change accordingly. \label{fig:map}}
\end{figure}

Neugebauer et al.~\cite{Neugebauer_2009_Eurovis} introduced a map
display for scalar flow features, where the 3D anatomy model is
presented in a central part and flow features of the surrounding sides are presented as flat regions of a map of the anatomical view. The map views and the 3D anatomy view are linked to depict positions of interest selected by the user (see
Figure~\ref{fig:map}).  This
enables a systematic exploration of all regions. Neuroradiologists
emphasized that this technique enables a better exploration of
scalar flow features.

Despite encouraging feedback obtained by Neugebauer et al., more evaluation and
corresponding refinements are necessary to make this strategy
broadly applicable. While in principle their approach is applicable to unsteady flows, it is likely that modifications are necessary if the
scalar flow features change over time, leading to frequent changes
of both views. Furthermore, flow information is volumetric. Although some measures are meaningful on the vessel wall, several flow features can only be analyzed through full volumetric visualization.

\subsection{Blood Flow Probing}

Time consuming segmentation is a necessary step in most tools for the inspection of measured flow data. Many tools incorporate functionality for local 2D segmentation of the vasculature~\cite{gyrotools:2011}. This step is necessary not just to provide the anatomical context, as described in section~\ref{section-anatomicalcontext}, but also to limit the domain where the vector field is considered valid. Due to the acquisition process, measured flow data presents values outside the vessel boundaries. This limits direct applicability of some visualization methods, such as integral lines or particle traces.

Segmentation techniques have been developed to directly segment measured 4D flow data. These methods are based on the assumption that the flow outside the vessel boundary exhibits incoherent behavior~\cite{chung:2004,solem:2004,persson:2005}. Van Pelt et al.~\cite{Pelt:2011b} presented an extension of active surfaces to segment flow. Krishnan et al.~\cite{Krishnan:2011} introduce a segmentation technique based on Finite-Time Lyaponov Exponents (FTLE). The main drawback of these methods is that they will fail in some pathologies due to the characteristics of the flow, e.g., areas with slow flow.

Most flow visualization techniques require seeding or region selection as initialization. The main reason for the selection is to avoid the clutter that visualizing the flow in the full domain supposes. The definition of the seeding region is usually done by probing in the volume domain, often with the help of segmentation.

Van Pelt et al.~\cite{Pelt:2010} presented a semi-automatic technique to probe cross-sections of anatomical data avoiding full segmentation ( see Figure~\ref{fig:vanPelt2010:context}). If anatomical data is not available an option for cross-sectional placement is to use the so called temporal maximum intensity projection (TMIP). For each voxel position of the TMIP scalar volume, the maximum speed is determined along the time axis of the 4D flow data. Hence, each voxel with a bright intensity indicates that a flow velocity with a substantial speed has occurred there at least once during the cardiac cycle. This probing method has several drawbacks: it assumes tubular structures, so it is only valid for vessels, and it does not consider the movement of the vessels during the heart cycle.

In later work, Van Pelt et al~\cite{Pelt:2011a} presented a probing technique to allow fast qualitative inspection, avoiding full segmentation. The user positions a 3D virtual probe on the viewing plane with common 2D interaction metaphors. An automatic fitting of the probe is provided for the third dimension, i.e., the viewing direction. Using the available velocity information of the measured blood flow field, the probe is aligned to this field. In particular, the automated fitting aligns the orientation of the long axis of the virtual probe to be tangential to the average local blood flow orientation. The probe is the basis for further visualizations (see Figure~\ref{fig:van Pelt2011:vis}).

Van Pelt et al.~\cite{Pelt:2010} also investigated different local seeding strategies based on the vessel center (e.g., radial or circular)  concluding that fixed template seeding cannot accommodate flow variations. Krishnan et al.~\cite{Krishnan:2011} presented a seeding strategy based on the segmentation of flow maps~\cite{Soni:2008}. Flow maps are based on the end position of the particle after integration or advection. It is expected that this seeding strategy will adapt to real flow patterns.

\begin{figure}[h]
\centering
 \subfigure[]{\includegraphics[height=37mm]{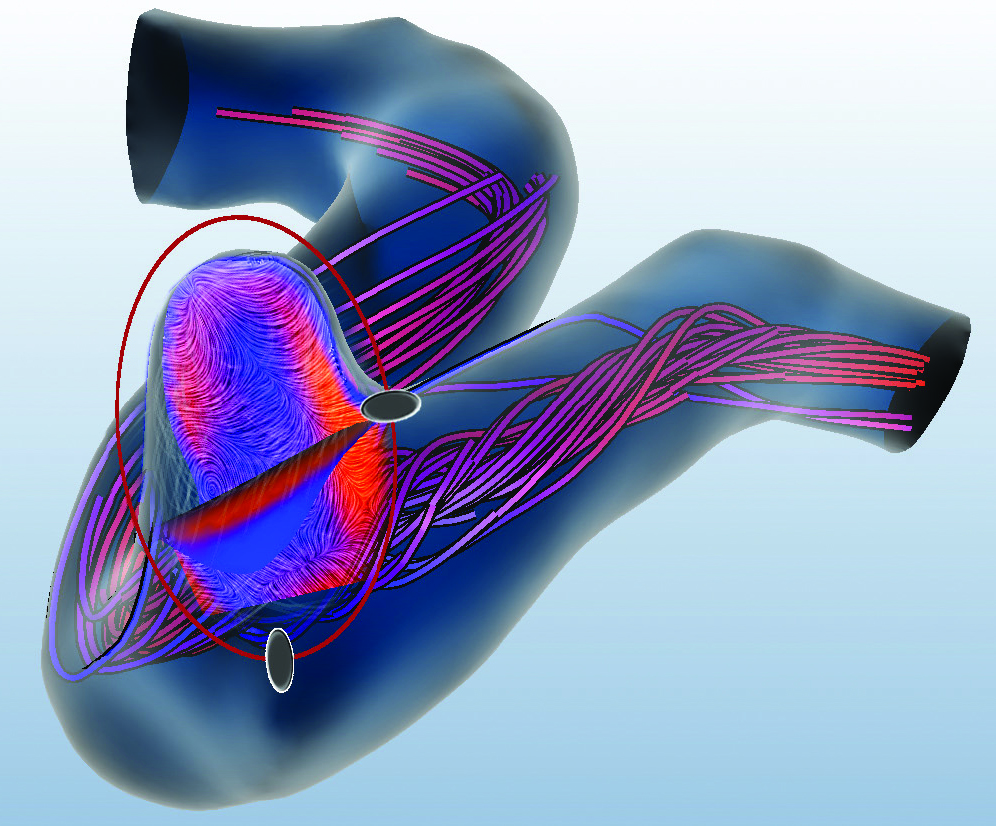}}\quad
 \subfigure[]{\includegraphics[height=37mm]{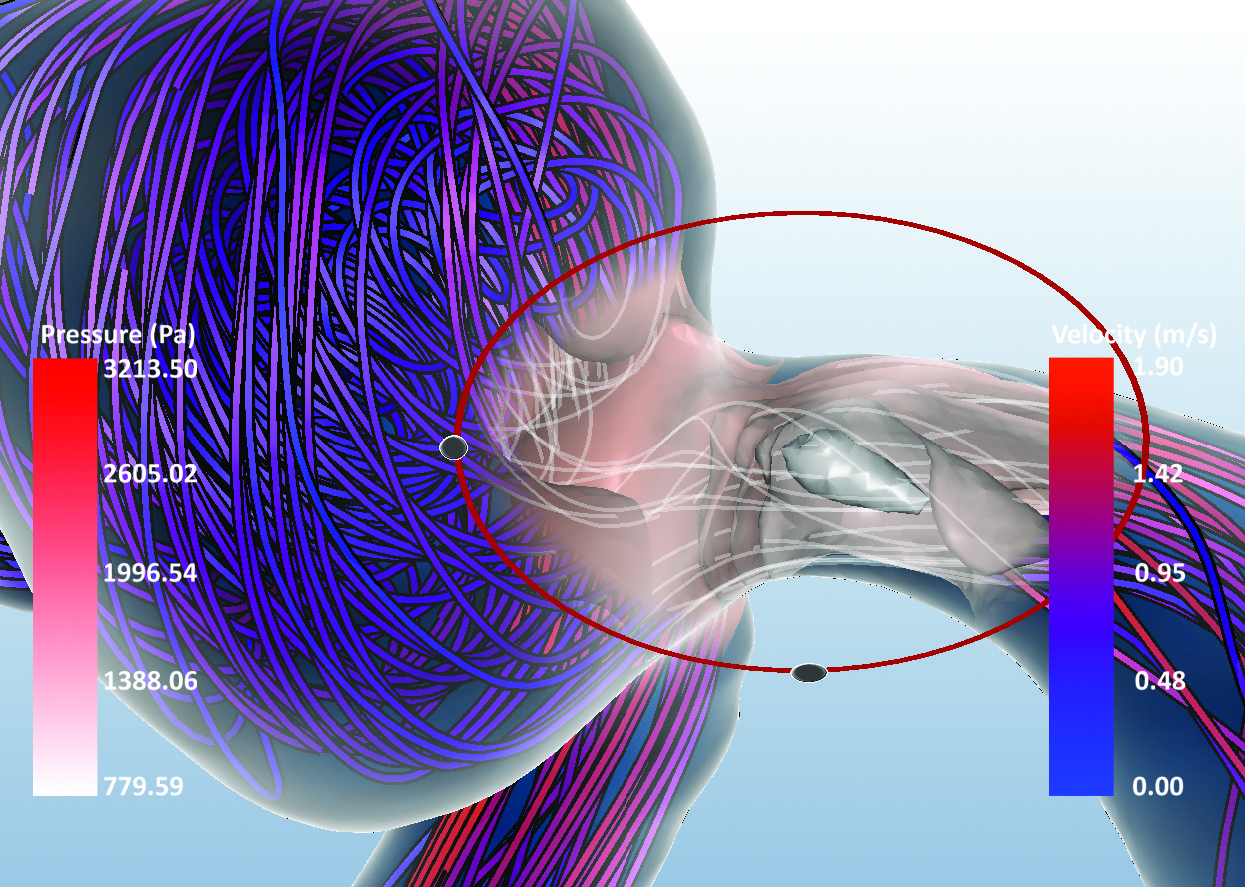}}
\caption{ Two examples of the FlowLens, which is depicted with a red contour line and two handles. Outside the lens the flow is visualized with illustrative and color-coded streamlines. Inside the lens a view-aligned probe plane with a LIC visualization for the investigation of the degree of vorticity (a) as well as the flow pressure as isosurfaces (b) are embedded. Additionally, the ostium surface is shown within the lens of the  LIC visualization as anatomical landmark.}~\label{fig:flowlens}
\end{figure}

In visualization, focus-and-context approaches are commonly used to avoid clutter. Gasteiger et al.~\cite{Gasteiger_2011_VIS} propose the FlowLens which is a user-defined 2D magic lens. This lens combines flow attributes by showing a different attribute and visualization within and outside the lens. Additionally, they incorporate a 2.5D lens to enable probing and slicing through the flow. To simplify the interface, they provide scopes which are task-based. Each scope consists of pairs of focus vs. context attributes, and propose visualization templates to represent each pair (see Figure~\ref{fig:flowlens}).

\subsection{Blood Flow Visualization}\label{section-flowvisualization}

Flow visualization has been an active field of research for decades. It has developed a large number of methods for inspecting flow data, and there are different review articles that define and classify these techniques~\cite{Post:2003,Laramee:2004}. Although these techniques can be directly applied to blood flow fields, it is important to note that not all techniques are meaningful due to the characteristics of the data, e.g., measured data has low temporal resolution. Furthermore, the chosen visualization should be comprehensible to physicians and clinical researchers. In other words, the features shown should be linked to an intuitive understanding of the flow, and the pathology. In the remainder of this section, we will present the most common blood flow visualization techniques that have been proposed in literature, and their variations.

One of the most common ways to depict flow is using \emph{integral curves}. There are two main approaches used for blood flow the so-called streamlines and pathlines. These integral curves represent the trajectory that a massless particle would follow through the vector field. Streamlines assume steady flow, so the vector field does not change in time. Pathlines are the extension of the streamlines that convey the temporal behavior of unsteady flow fields. Therefore, pathlines are the curves to depict particles trajectories in the vascular system. However, streamlines are still often used to depict instantaneous flow-field structure. In measured flow data, where the temporal resolution is low, streamlines can be informative, since error is accumulated at each integration step of the pathlines, and therefore the reliability of the lines decreases rapidly. Streaklines are another category of integral curves which have been used less often for blood flow. Streaklines are generated by a continuous seeding through time. Each point of the line corresponds to a seed that is continuously integrated through time.

\begin{figure}[h]
\centering
\subfigure[]{\includegraphics[height=37mm]{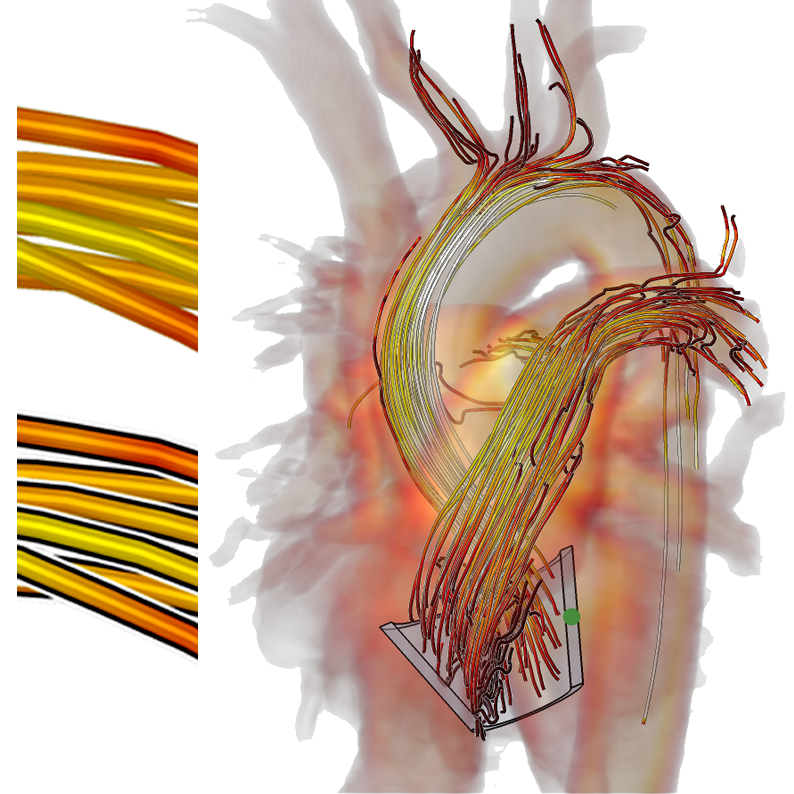}}\quad
\subfigure[]{\includegraphics[height=37mm]{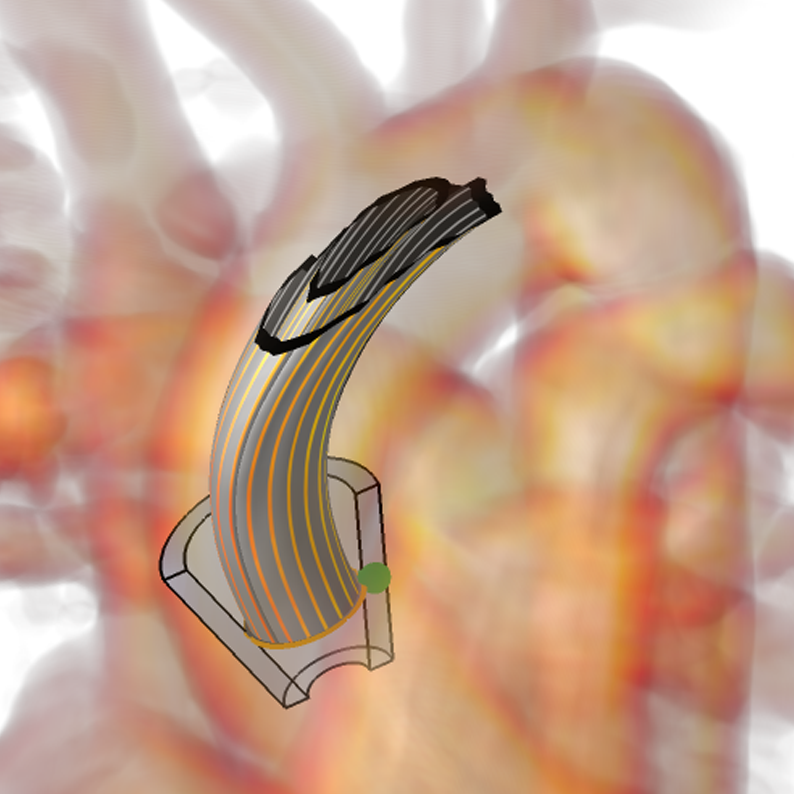}}\quad
\subfigure[]{\includegraphics[height=37mm]{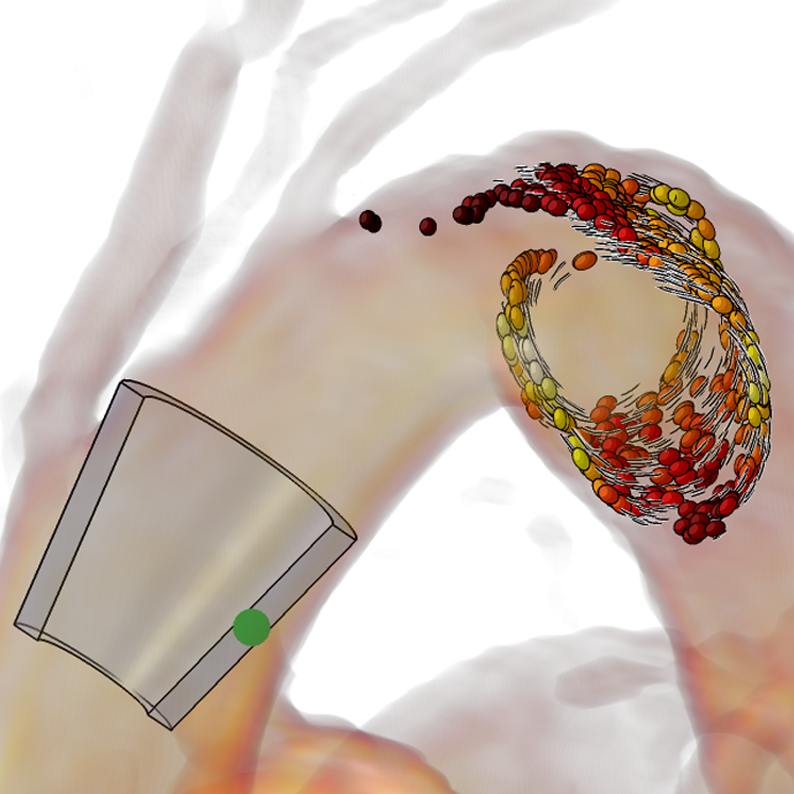}}
 \caption{GPU-generated long pathlines, statically seeded in the heart chambers to capture aorta and pulmonary arteries. (a) shaded imposter tuboids. The left-hand side shows a larger versions of the pathlines, respectively without and with halos and contours. (b) pathsurfaces presented as tube-shaped surfaces. The latter may be nested, approximately capturing the wave-front profile. A stripe pattern is employed to convey rotation~\cite{Pelt:2011a}~\label{fig:van Pelt2011:vis}}
\end{figure}

Integral curves are often rendered as illuminated lines or shaded tuboids. Perception of the spatial relations between pathlines is improved by means of halos~\cite{everts:2009}. Van Pelt et al.~\cite{Pelt:2011a} additionally applied contours in order to enhance the structure of the pathlines (see Figure~\ref{fig:van Pelt2011:vis}(a)).
Two types of seeding strategies are also proposed by Van Pelt et al~\cite{Pelt:2010}. On the one hand, lines may be seeded statically from a fixed position in space and time. On the other hand, integration curves may be seeded dynamically, tracing the lines from a fixed spatial location, and varying seed time with the current time frame of the cardiac cycle. Dynamically seeded pathlines consist of comparatively short traces. Although the covered temporal range is relatively narrow, the pathlines are more reliable, and provide an approximative depiction of the pulse-wave in the cardiovascular system. The drawback is that it does not provide enough information on a large scale; it only provides sufficient local information.


\emph{Integral surfaces} are a generalization of integral lines. Integral surfaces are formed by a continuum of integral curves. The surfaces enable surface shading techniques which improve the perception of 3D structures. Integral surfaces are initialized by a seeding curve which defines the topological connection between the integral curves. Depending on the integral curve used there exist: streamsurfaces, pathsurfaces and streaksurface. Integral surfaces have been recently studied for blood flow~\cite{Krishnan:2011,Pelt:2011a}.

The seeding curve used for initialization is crucial for the correct interpretation of the integral surface. Krishnan et al.~\cite{Krishnan:2011} define the seeding curve as the boundary of segmented regions based on flow maps. Van Pelt et al.~\cite{Pelt:2011a} presented cross-shaped and tube-shaped patterns of the integral surfaces (see Figure~\ref{fig:van Pelt2011:vis}(b)). Integral surfaces allow shading and texturing. For example, for tube-shaped surfaces stripes texturing emphasizes the rotational motion around the centerline. The color may convey various derived measures of the blood flow. In the user evaluation of Van Pelt et al.~\cite{Pelt:2011a}, the integral surfaces were considered valuable to explore the local rotational aspects of the flow.

\emph{Particle systems} are readily applied to blood flow. Commonly, particles are depicted as spheres~\cite{nvidia:2010}, or otherwise represented by small integral curves~\cite{gyrotools:2011}. Both approaches convey blood flow speed through color, while direction information is captured by temporal cohesion of the animated particles. Integral curves additionally provide a short history of the particle trajectory. Both conventional approaches employ the available visual cues, such as color and shape, to capture only the blood flow velocity information. Van Pelt et al.~\cite{Pelt:2011a} propose an illustrative particle-based approach that captures the velocity information by means of shape, keeping the color cue available for more elaborate blood flow characteristics. They mimic techniques often used to convey motion in comic books by deforming a ball at high-speed motion, and adding speed lines to improve the perception of direction (see Figure~\ref{fig:van Pelt2011:vis}(c)).

Visual clutter remains an important issue in 4D blood flow field visualizations. Usually, this clutter is avoided by user biased interaction methods which can miss important properties of the flow. Grouping vector field areas with meaningful similar characteristics, i.e., clustering, can help in developing techniques to improve the visual exploration and minimize the user bias. Some work exists in the clustering of static 3D vector fields~\cite{Telea:1999,Garcke:2000,Li:2006,Kuhn:2011} while little research has been conducted to extend it to 3D unsteady flow fields~\cite{Yu:2007}. Developing and extending these techniques to blood flow data is an interesting research direction. The main challenge is to provide a clustering that has a meaning for the user, and an adequate visualization technique that enables efficient exploration of the clusters.

\section{Discussion and Open Issues}\label{section-discussion}

Simulated and measured blood flow data have been two distinct research fields that have developed in parallel. Simulation data is based on many assumptions, it is difficult to make it patient-specific, and also validation is a challenge. Measured flow data, on the other hand, represents the patient-specific flow, but it has a lot of limitation concerning resolution, artifacts, and noise in the data. An interesting direction is to combine both methods to strengthen each other. For example, measured data can be used as boundary conditions for a simulation, or simulation methods could be used to compensate for the lack of temporal resolution. 

Blood flow data sets are considerably large data sets since they consist on a time series of vector-field volumes. The issue of dealing with large data will be of increasing importance given the improvements in spatial and/or temporal resolution that are expected.

Recently, new blood flow visualization techniques have been developed. Many decisions with respect to seeding, segmentation, integral curves, the use of illustration techniques are rather ad-hoc decisions based on intuition. It is important to link the decision to the users' needs. Additionally a more thorough exploration of the design space and comparisons of existing methods is needed. A major challenge is that this data is new to the domain experts, and it is challenging for them to identify the relevant features to visualize.

The existing methods are rather complex visual representations that overwhelm a considerable portion of the target user group. Future research should address simplifications of the blood flow by either clustering flow or by detecting and emphasizing relevant features. Existing techniques for flow field analysis may serve as orientation, but certainly need to be combined with the in-depth knowledge of domain experts regarding the relevance of certain blood flow characteristics. A better understanding of specific tasks, decisions and relevant information is necessary to support blood flow exploration with a guided workflow-based interaction.

Radiologists need to prepare reports where they summarize their findings verbally including relevant images. A better understanding of such reports may help to better support reporting, e.g., in case of cardiovascular diseases. While current applications are strongly focused on measured cardiac flow and simulated cerebral blood flow, advances in image acquisition will lead to further applications, e.g., where renal or liver flow is represented.


\bibliographystyle{splncs}
\bibliography{bibtex/bloodFlowBernhard,bibtex/Annavilanova}

\end{document}